# A comparative study of scheduling techniques for multimedia applications on SIMD pipelines

Mehmet Ali Arslan, Flavius Gruian and Krzysztof Kuchcinski
Department of Computer Science, Lund University, Sweden
Email: {mehmet_ali.arslan, flavius.gruian, krzysztof.kuchcinski}@cs.lth.se

*Abstract*—Parallel architectures are essential in order to take advantage of the parallelism inherent in streaming applications. One particular branch of these employ hardware SIMD pipelines. In this paper, we analyse several scheduling techniques, namely ad hoc overlapped execution, modulo scheduling and modulo scheduling with unrolling, all of which aim to efficiently utilize the special architecture design. Our investigation focuses on improving throughput while analysing other metrics that are important for streaming applications, such as register pressure, buffer sizes and code size. Through experiments conducted on several media benchmarks, we present and discuss trade-offs involved when selecting any one of these scheduling techniques.

## I. INTRODUCTION

With the growing demands on mobile computing, fuelled among others by the advent of the Internet of Things (IoT) and cloud computing, the amount of processing required from multimedia and telecommunication applications are also increasing. Such applications typically process streams of data, exhibiting a high degree of parallelism, requiring high performance with a limited power budget. The answer, from the hardware point of view, is the introduction of highly parallel architectures.

Parallelism can be achieved at different levels in an architecture, by employing pipelining, SIMD processing, multi-core or array processors. The current trend in commercial system is to use multi-core systems extended with graphic processors (GPU). In mobile devices the alternative is often offered by using DSP processors and ASIC accelerators. Furthermore, some applications require custom architectures, designed especially to provide high computational power, for very specific programming models.

Combining all these techniques and parallelism at different levels can in principle yield architectures with enough computational power and low power consumption, but their programming is not trivial. The main challenge today is therefore programming these platforms such that the applications fully utilize the hardware resources. Furthermore, different implementation choices offer different trade-offs in terms of performance, memory consumption, power, etc. Therefore, depending on the application context, it is important to carefully consider alternatives rather than committing to the "best" implementation.

In this paper, we analyse and compare different ways of scheduling repetitive behavior (kernels) when compiling code for parallel architectures. In particular, the target architecture we focus on is a generic architecture that employs a SIMD pipeline. This architecture model is abstract enough to model a class of architectures, including a custom reconfigurable architecture designed for our specific application area [1]. That architecture centres around a highly reconfigurable pipelined processor with vector instructions (SIMD). Reconfigurations may be carried out in one clock cycle, which makes for a very flexible instruction set, offering a large number of execution and optimization choices. Furthermore, the architecture model we adopt in this paper is also rather similar to the GPU hardware.

The remainder of the paper is organized as follows. Section II briefly presents the previously published work in the context of scheduling for pipelines and SIMD architectures. Section III presents the generic architecture model as well as other assumptions adopted in the paper. Section IV describes the scheduling techniques studied in this paper along with a brief overview of the method we used to obtain such schedules. Section V presents the experimental setup and results while section VI employs these results to compare the techniques described earlier. Finally, our conclusions are drawn in section VII.

## II. RELATED WORK

Optimal use of the instruction level parallelism (ILP) through software pipelining [2] has been addressed in the literature early on. One of the standard techniques, first proposed in [3], is *modulo scheduling*, which selects a minimal schedule for one loop iteration such that no constraints are violated when the schedule repeats.

Combined with retiming and unrolling, modulo scheduling can be even more effective, as shown in [4], [5], [6]. Nevertheless, such techniques, intended to increase the instruction per clock (IPC) count or throughput, suffer instead from a drastic increase in live data, thus register requirements. Therefore, much of the work also focused on minimizing the register pressure [7], [8], [9] while scheduling.

Computing architectures with high ILP that benefit from software pipelining are today ubiquitous. One of the earliest uses of the technique is described in [10] which targets very large instruction word (VLIW) architectures. Vectorization for single instruction, multiple data (SIMD) architectures has also been tackled early on [11], [12], [13], while more recent work even targets streaming applications [14]. Software pipelining and especially modulo scheduling was also proven to be successful for newer architectures, designed for offering a high degree of parallelism, namely coarse-grained reconfigurable arrays (CGRA) [15], [16], [17]. Even more exotic architectures, especially designed for streaming applications have been shown to benefit from software pipelining [1], [18].

In this paper we revisit modulo scheduling along with partially unrolled modulo scheduling, as well as another common but ad hoc practice we call *overlapped execution* (see section IV-B). However, the focus of our investigation is on those measures that are relevant for streaming applications, namely throughput and input/output data rates. Register pressure (minimal register requirements) as well as code size are also examined, in order to underline the differences between the various techniques. The interaction between registers, extent of unrolling and code size has been studied previously in [19]. In that work, however, the focus is on VLIW processors, and unrolling is basically carried out after scheduling, while we target SIMD processors, and unroll before scheduling.

For streaming applications, using modulo scheduling in order to increase throughput, while also keeping buffer sizes under control, has already been addressed in the literature [20], [21], [22]. However, it is important to notice that the type of pipelining targeted in these approaches is *algorithmic pipelining* on multi/many-cores. In that case scheduling employs processor level parallelism (PLP), rather than ILP as in our case. By replicating tasks (*actors*) and groups of actors on the same or several processors, data production and consumption rates are negotiated in such a way that buffers sizes are minimized. Note that each instance of an actor execution is treated as an atomic behaviour with a given latency. The type of scheduling we address in this paper is complementary to the algorithmic pipelining used in those approaches, adding another degree of freedom to the design flow. Instead of assuming a fixed atomic execution (yet repetitive) of an actor, we show that its behaviour can be pipelined in different ways, yielding different input/output rate which will affect the choices of algorithmic pipelining.

An approach that does use both ILP and PLP for streaming application is described in [14], employing SIMDization at several levels. However, their use of ILP is restricted to stateless actors, whose instances are combined to exactly cover the vector size and execute in parallel. A similar technique for loop SIMDization, focused on memory access patterns, is presented in [13]. In contrast, our use of ILP allows for more generic architectures, execution models, and actors.

## III. CONTEXT

As target applications, we consider kernels from digital signal processing (DSP), which are part of image processing algorithms. Such kernels can be implemented differently depending of the computational model. For example, in an imperative language they are usually implemented as an iterative execution of a code sequence (a loop or a nested loop construct). In the dataflow model of computation [23], this is equivalent to the repetitive execution of an actor processing a data stream.

For this study, we confine ourselves to kernels that have no inter-iteration dependency, which allows us to emphasize the differences between various execution scenarios. However, our scheduling methods are not limited by this assumption and can be extended with additional constraints modelling inter-iteration dependencies.

In the following, we use benchmarks that represent inner loop bodies or actor computations. Each benchmark is modelled as a *directed acyclic graph* (DAG) whose nodes represent basic operations and edges dependencies between them. Similar basic operations, which are also independent, can thus be grouped together and issued as one SIMD instruction.

We consider a generic target architecture with a hardware pipeline that can perform SIMD operations of a given width. The hardware pipeline has a given number of stages, which defines the latency between data dependent operations, since we assume that data must be available at pipeline start and produced by the last stage. Furthermore, we assume that each pipeline stage has a latency of one clock cycle. This particular choice of the architecture is motivated by our previous work with the system in [1].

## IV. APPROACH

We compare different scheduling techniques for DSP kernels, usually employed to increase the throughput. The measures we examine for each technique are those we consider relevant for streaming applications, namely throughput, register pressure, input/output data rates, and code size. We also investigate how different architecture configurations influence these measures.

The techniques we consider are: scheduling a single iteration, overlapping iterations, along with modulo scheduling, classic as well as combined with loop unrolling. The experiments are carried out on several graphs from ExpressDFG [24] which provides dataflow graphs for benchmark applications from the MediaBench suite [25].

Practically the tools implementing the different scheduling methods were developed using the constraint programming environment offered by JaCoP [26]. The modelling of these methods using constraint programming is itself an interesting problem, which we addressed in detail in our previous work. In the following, without going into the modelling issues, we briefly present the scheduling techniques we investigate in this paper.

### A. Scheduling one iteration

The straightforward way to schedule a loop/kernel is to sequentially execute iterations, which requires scheduling one iteration efficiently. To illustrate, we will use a small part of an elliptic wave filter (EWF), taken from ExpressDFG [24]. Figure 1 shows the shortest schedule, given the assumptions described in section III.

In the figure, the DAG format is preserved to show the dependencies and schedule time is shown on the horizontal axis. We deliberately picked this part from EWF to illustrate the scheduling behavior when the available ILP is limited and the critical path dominates the schedule. In this example, the resource utilization, calculated by the number of operations that are scheduled, over the number of nodes the architecture can run during the schedule length, is around 6%. Throughput is 0.024 samples per clock cycle and latency is 42 clock cycles. This poor utilization is caused both by the properties of the architecture and the application, as we detail in the following.

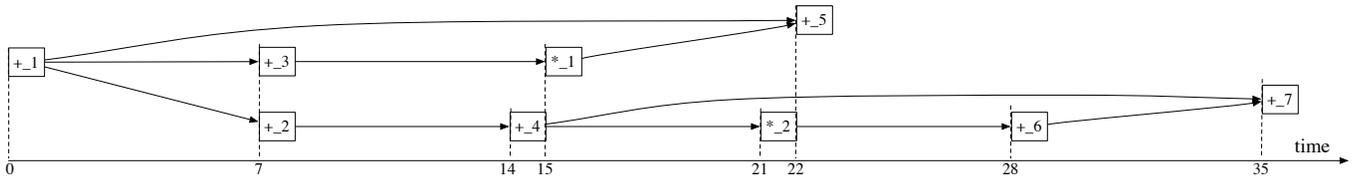

Fig. 1: Example schedule of a single iteration.

The long latency of the hardware pipeline (which is equal to 7 for this example - see section III) is one reason. Due to the data dependencies (mainly through the critical path), each dependent node has to wait for its predecessor to finish its execution, i.e. go through all the pipeline stages.

The low ILP present in the application is the other reason for the poor resource utilization. Note from the figure that some of the nodes are scheduled simultaneously. This is enabled by the SIMD nature of the architecture which execute up to SIMD width (4 in this example) number of operations. However, in this case there are not enough operations that are mutually independent, hence not many can be scheduled together.

Poor utilization generally means poor throughput and low energy efficiency. To increase both, it is common procedure to schedule several iterations simultaneously.

### B. Overlapping execution

For architectures with a hardware pipeline, an easy way to increase utilization is to schedule multiple iterations in an ad hoc, overlapping fashion. The scheduling process is two-fold. First the instructions for a single iteration are selected and ordered, with the objective of minimizing the number of effective (non-*nop*) instructions (in contrast to schedule length). The overlapped schedule is obtained then by advancing the chosen number of iterations by running the same corresponding instruction from each one in sequence. Once all equivalent instructions from all iterations have been scheduled, the execution advances to the next instruction. The resulting schedule for the example from Fig. 1 is given in Fig. 2, where iterations are denoted with upper case letters in the node identifier.

This computationally simple solution eliminates the gaps in the schedule caused by the hardware pipeline completely, as long as the number of iterations scheduled simultaneously is larger than the pipeline latency/length. For this example the utilization is increased to 32%. The throughput is 0.146 samples per clock cycle and latency is 42 clock cycles.

Besides its simplicity, this approach is suitable particularly for certain reconfigurable architectures for which the reconfiguration overhead is a significant issue in scheduling. A reconfiguration is needed when two different types of instructions follow each other, and this approach limits it to

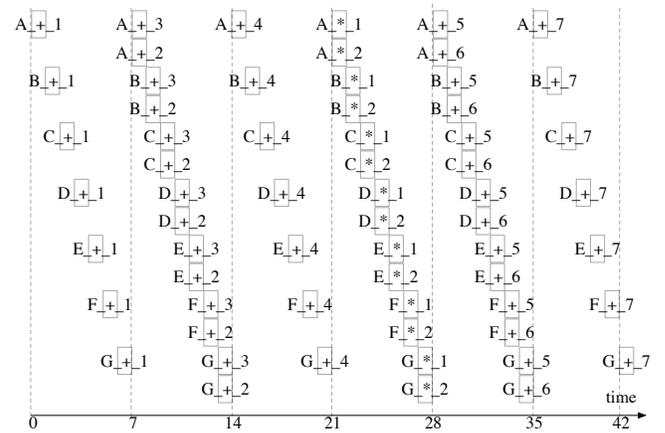

Fig. 2: Example schedule of an overlapped execution.

the number of instructions. (More details for this approach and its usage can be found in [27])

### C. Modulo scheduling

As mentioned in section II, *modulo scheduling* is a common technique for increasing the throughput of a kernel. It involves finding a schedule that initiates iterations as soon as possible, taking into account dependencies and resource constraints, while also repeating regularly with a given interval (*initiation interval II* [28]).

Scheduling the example from Fig. 1 gives an *II* of length 3, as depicted in Fig. 3a. Nodes from the same iteration are coded with the same upper case letter, in the same order as the iterations, to give an idea about how the *II* is assembled from different iterations. With this notation, node $A\_+\_7$ is the last operation from iteration $A$, while node $M\_+\_1$ is the first operation from the latest iteration about to start executing. Note that iteration $M$ is twelve iterations later than $A$. The iterations in between $B$ and $L$ have already started in previous *II*s and are currently executing (are active) in the pipeline. It is interesting to note however that not all iterations have operations that start in every *II* instance, although they may be in the process of executing operations or simply wait for data. With our notation, iterations $B, D, G, J, L$, which correspond to iterations with distances of $1, 3, 6, 9, 11$ relative to $A$, are *invisible* since they are not starting any operations in the *II* depicted in Fig. 3a. Note however that from an absolute point of view, the set of *invisible* iterations will change with the *II* instances. The existence of these iterations is an artefact of the short *II* relative to the rather long hardware pipeline.

For the given SIMD width of four, only the first clock cycle in Fig. 3a utilizes the hardware fully. Elsewhere the hardware is underutilized because of insufficient ILP in the application, even using modulo scheduling. The average utilization in the *II* is 75%, while the throughput is 0.33 samples per clock cycle and latency is 44 clock cycles.

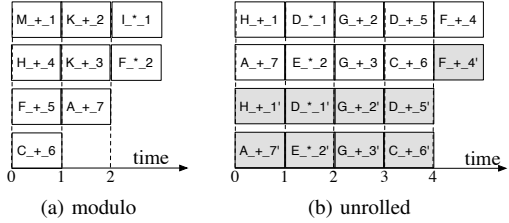

(a) modulo  (b) unrolled

Fig. 3: Initiation intervals for modulo and unrolled modulo scheduling, for the example in Fig. 1.

Compared to the overlapped execution, modulo scheduling increases the ILP, to some extent, both by folding the graph as well as filling in the gaps resulted from the pipeline latency.

### D. Unrolling and modulo scheduling

Even though modulo scheduling increases the available ILP by folding several iterations to construct the *II*, it can be seen from Fig. 3a that there are still slots in the schedule that are not utilized. To remedy this, an idea is to unroll several iterations prior to modulo scheduling. This way, the available ILP can be increased further.

In Fig. 3b we depict the *II* for our running example when two iterations are unrolled before modulo scheduling. In addition to the notation from Fig. 3a, we use different shading to distinguish between nodes unrolled from different iterations. For example, $H\_+\_1$ and $H\_+\_1'$ refer to the first node from the first and the second unrolled instances, respectively, which together constitute iteration H. With the increased number of independent nodes, this method makes better use of the SIMD capability compared to classic modulo scheduling. Thus the utilization increases to 90% for this example, throughput is increased to 0.40 samples per clock cycle while the latency becomes 42 clock cycles.

## V. EXPERIMENTS

In this section, we present our experiments for measuring the quality of the schedules provided by the methods introduced earlier. A deeper discussion around these results makes the subject of the following section. For these experiments, unless stated otherwise, we assumed that target architecture, introduced in section III, has a pipeline length of seven stages, and a SIMD width of four. These parameters are varied occasionally as stated, in order to investigate their impact on the measures under scrutiny.

To illustrate various characteristics of the scheduling techniques, we use Loefller's IDCT [29] (referred to as *IDCT* in the rest of the paper) to carry out a more detailed analysis. IDCT is part of many image and video applications including JPEG and MPEG, and Loeffler's IDCT is a commonly used implementation. Note however, that similar results are observed for the other kernels we experimented with (see Tab. II).

| IDCT ($|V|=48, |E|=63$) | Single | Overlapped | Modulo | Unrolled x2 |
|---|---|---|---|---|
| Throughput (samples/cc) | 0.022 | 0.046 | 0.059 | 0.080 |
| Latency (cc) | 46 | 147 | 53 | 67, 68 |
| Code size (instr.) | 46 | 153 | 17 | 25 |

TABLE I: Loeffler's IDCT with various scheduling techniques.

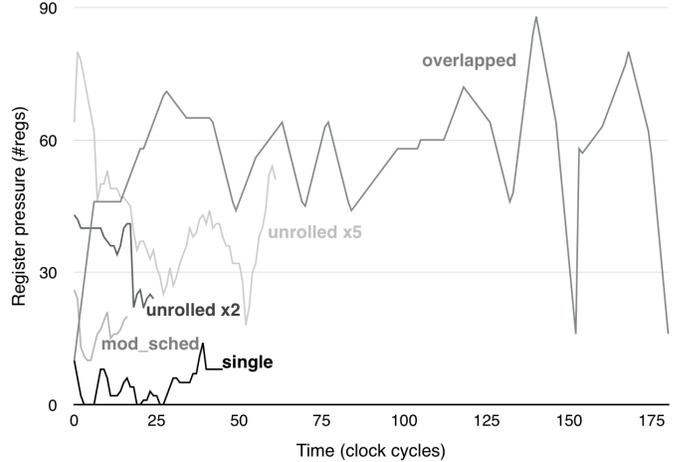

Fig. 4: Register pressures for IDCT scheduled with various techniques. Each plot covers one execution of the respective schedule.

A summary of the results for scheduling IDCT is given in Tab. I. Number of nodes and edges in the graph are denoted with $|V|$ and $|E|$, respectively. Each column refers to a different scheduling technique. The column "single" refers to the results for scheduling one iteration and constitutes a baseline for comparison. For data related metrics, one iteration is assumed to consume and produce one unit of data, referred to as a *sample* in the table. Overlapped scheduling interleaves seven iterations, since the pipeline length/latency is seven. Unrolled modulo scheduling, referred to as *unrolled* in the rest, involves unrolling two iterations.

For single and overlapped scheduling, the *throughput* is calculated as #*iterations*/*schedulelength*. When computing this for modulo and unrolled the *schedulelength* is replaced by *II*, since at the end of each *II*, exactly #*iterations* finish executing and produce their outputs. The *latency* for an iteration is the distance between its first scheduled input node and last scheduled output node. The latency of all iterations in an overlapped schedule is the same since each iteration is only a shifted version of the first. In contrast, as the unrolled iterations are scheduled freely to obtain a minimal *II*, their latencies may differ.

Code size denotes the number of instructions that would be generated from the schedule. This includes the possible no operation (*nop*) instructions. For single and overlapped scheduling this figure is equal to the schedule length. For modulo and unrolled, the prologue and epilogue are negligible when the kernel is run many times, thus the code size for these methods is equal to their *II*.

Assuming that live data is kept in registers, the number of registers needed throughout each schedule is plotted in Fig. 4. Computing the lifetimes for input/output data is interesting, since this is directly coupled to the buffer sizes, when it comes to streaming applications. Input data is considered alive from the first operation using (part of) it, while output data is alive until the last operation producing (part of) it. Input data becoming alive is reflected through the consumption of an input sample from the input buffers. Similarly for output data end of life and sample production in the output buffers.

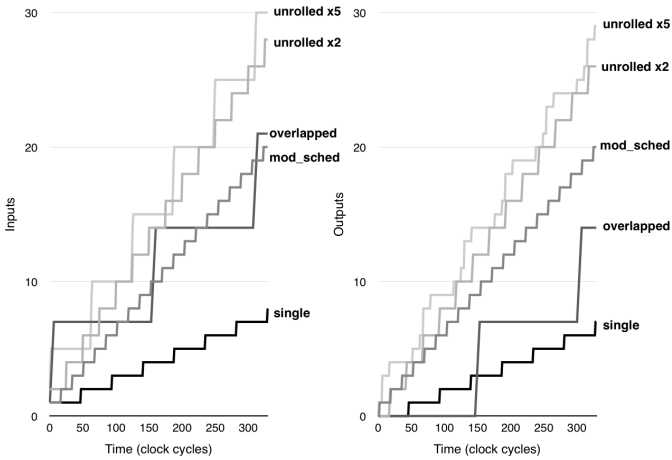

Fig. 5: Input consumption and output production for IDCT scheduled with various techniques.

| EWF ($|V|=34, |E|=47$) | Single | Overlapped | Modulo | Unrolled x2 |
|---|---|---|---|---|
| Throughput (samples/cc) | 0.010 | 0.067 | 0.111 | 0.118 |
| Latency (cc) | 98 | 98 | 104 | (98, 98) |
| max. reg. press (registers) | 9 | 41 | 49 | 52 |
| data cons. (samples) | 1 | 7 | 1 | 1 |
| data prod. (samples) | 1 | 7 | 1 | 1 |
| Code size (instr.) | 98 | 104 | 9 | 17 |
| **JPEG FDCT** ($|V|=134, |E|=169$) | **Single** | **Overlapped** | **Modulo** | **Unrolled x2** |
| Throughput (samples/cc) | 0.011 | 0.020 | 0.025 | 0.026 |
| Latency (cc) | 92 | 343 | 122 | (157, 155) |
| max. reg. press (registers) | 26 | 198 | 64 | 62 |
| data cons. (samples) | 1 | 7 | 1 | 2 |
| data prod. (samples) | 1 | 7 | 1 | 1 |
| Code size (instr.) | 92 | 349 | 40 | 78 |
| **MPEG IDCT** ($|V|=114, |E|=164$) | **Single** | **Overlapped** | **Modulo** | **Unrolled x2** |
| Throughput (samples/cc) | 0.009 | 0.016 | 0.021 | 0.023 |
| Latency (cc) | 115 | 420 | 138 | (143, 143) |
| max. reg. press (registers) | 26 | 204 | 48 | 68 |
| data cons. (samples) | 1 | 7 | 1 | 1 |
| data prod. (samples) | 1 | 7 | 1 | 1 |
| Code size (instr.) | 115 | 426 | 48 | 88 |

TABLE II: Performance measures for three benchmarks scheduled with various techniques.

| # iterations unrolled | two | three | four | five |
|---|---|---|---|---|
| Throughput (samples/cc) | 0.080 | 0.075 | 0.083 | 0.081 |
| Utilization (n/(II*N))(%) | 96 | 90 | 100 | 97 |
| Latencies (cc) | 67,68 | 79, 85, 82 | 52, 97, 96, 91 | 63, 68, 78, 113, 66 |
| max. reg. press (registers) | 43 | 57 | 66 | 80 |
| data cons. (samples) | 1 | 2 | 2 | 3 |
| data prod. (samples) | 2 | 3 | 2 | 2 |
| Code size (instr.) | 25 | 40 | 48 | 62 |

TABLE III: The effect of changing the number of iterations unrolled for IDCT.

| SIMD width = 8 | Single | Overlapped | Modulo | Unrolled x2 |
|---|---|---|---|---|
| Throughput (samples/cc) | 0.022 | 0.053 | 0.143 | 0.143 |
| Latency (cc) | 46 | 126 | 53 | 57, 57 |
| max. reg. press (registers) | 12 | 88 | 47 | 61 |
| data cons. (samples) | 1 | 7 | 1 | 1 |
| data prod. (samples) | 1 | 7 | 1 | 1 |
| Code size (instr.) | 46 | 132 | 7 | 14 |
| **SIMD width = 16** | **Single** | **Overlapped** | **Modulo** | **Unrolled x2** |
| Throughput (samples/cc) | 0.022 | 0.072 | 0.250 | 0.286 |
| Latency (cc) | 46 | 97 | 52 | 53, 53 |
| max. reg. press (registers) | 14 | 88 | 73 | 93 |
| data cons. (samples) | 1 | 7 | 2 | 1 |
| data prod. (samples) | 1 | 7 | 2 | 1 |
| Code size (instr.) | 46 | 97 | 4 | 7 |

TABLE IV: The effect of SIMD width on IDCT schedules.

Note again, that the overlapped schedule comprises seven iterations. Modulo and unrolled schedules are plotted only over one *II*, since this profile repeats every *II*. Here *unrolled x2* unrolls two iterations and *unrolled x5* unrolls five.

Throughput is a performance metric that often refers to the average output of the schedule. However, for some application domains, such as streaming, the instant rates of input consumption and output production is more important, also affecting the buffer sizes. To show the variation in these rates between the various scheduling techniques, Fig. 5 plots the cumulative input consumption and output production for IDCT, over a longer time period.

Tab. II summarizes the results for scheduling three additional kernels, taken from ExpressDFG [24]. Number of nodes and edges for each graph is denoted with $|V|$ and $|E|$, respectively, under the application name. The last kernel in Tab. II is a different IDCT implemented in MPEG, which should not be confused with Loeffler's IDCT.

Apart from the metrics in Tab. I, there are three additional metrics that summarize the register pressure, input and output rates for these kernels. (For IDCT, Fig. 4 and 5 present this information in more detail). The metric "max. reg. press" denotes the maximum register pressure in the schedule. To present the input/output behavior, we also use two other metrics, namely "data cons." and "data prod." Both use a time window that slides through the schedule and captures the maximum sample consumption and production, respectively. The length of this sliding window is set to the number of pipeline stages, in order to better capture the behavior of the overlapped schedule.

In order to show the differences between different extents of unrolling, we include *unrolled x5* and *unrolled x2* in both Fig. 4 and Fig. 5.

Further results from experiments with different extent of unrolling are given in Tab. III. In addition to previously mentioned metrics, we also include resource utilization. This is computed as $n/(II*N)$ where *n* denotes the number of nodes in the graph and *N* denotes the SIMD width (i.e. the number of operations that can be run simultaneously). We include utilization figures in order to give an idea of the efficiency of the schedule and the untapped parallelism still existent in the architecture. The "Latencies" row shows that the different iterations unrolled and scheduled together may in fact get different latencies. The reason is that latency is not in any way constrained in our CP model for modulo scheduling. If the application domain requires the latency for each sample output to be equal (or close to equal) this can be included in the model. However, it should be noted that the resulting *II* can degrade, since the solution space becomes more constrained.

Besides the target application and the scheduling method, there are two parameters of the architecture that can affect the schedules, namely the SIMD width and the number of pipeline stages. We varied both parameters and gathered the results for IDCT in Tab. IV and V, respectively.

| # Pipe. Stages = 4 | Single | Overlapped | Modulo | Unrolled x2 |
|---|---|---|---|---|
| Throughput (samples/cc) | 0.037 | 0.051 | 0.067 | 0.080 |
| Latency (cc) | 27 | 76 | 30 | 49, 49 |
| max. reg. press (registers) | 16 | 52 | 18 | 35 |
| data cons. (samples) | 1 | 4 | 1 | 1 |
| data prod. (samples) | 1 | 4 | 1 | 1 |
| Code size (instr.) | 27 | 79 | 15 | 25 |
| # Pipe. Stages = 16 | Single | Overlapped | Modulo | Unrolled x2 |
| Throughput (samples/cc) | 0.010 | 0.046 | 0.071 | 0.077 |
| Latency (cc) | 100 | 336 | 102 | 113, 115 |
| max. reg. press (registers) | 14 | 196 | 35 | 48 |
| data cons. (samples) | 1 | 16 | 2 | 1 |
| data prod. (samples) | 1 | 16 | 2 | 2 |
| Code size (instr.) | 100 | 351 | 14 | 26 |

TABLE V: The effect of pipeline length on IDCT schedules.

## VI. DISCUSSION

It is obvious from the previous section that the different scheduling techniques offer different trade-offs between the performance metrics we consider. Depending on the context, the choice of one or another technique may be preferred. However, we can claim with high confidence that scheduling a single iteration results in overall poor performance.

At a first glance the overlapped execution offers advantages only over the single iteration schedule, in terms of throughput. However, there are some hidden advantages that should be considered. First, obtaining an overlapped schedule is easier than obtaining modulo and unrolled schedules, which require specific techniques. In fact starting from a single iteration schedule, obtaining the overlapped schedule is straightforward. Second, overlapped execution may in fact be more effective on selected reconfigurable architectures, designed for streaming applications [27]. To keep our architecture abstraction simple, we did not include reconfigurability in this work. However in reconfigurable architectures where reconfiguration takes a certain time (e.g. one clock cycle) to switch between different instructions, overlapped can outperform modulo scheduling in terms of throughput. We present further insights into the effects of reconfiguration for overlapped and modulo scheduling in [27].

### A. Average throughput

In all examples, modulo and unrolled provide significant improvement in throughput over the overlapped scheduling, which is itself an improvement from single iteration scheduling.

The number of iterations that are unrolled affects the throughput as well. This is captured in Tab. III. For IDCT, unrolling four iterations gives the best throughput since it reaches full utilization. Further unrolling is in principle not needed and introduces mismatches between the application ILP and the architecture. Obviously for a different application, the optimum number of unrolls will be different. It should be noted as well, that increasing the number of unrolled iterations increases in turn register pressure and code size, and possibly also the required buffer sizes. Additionally it may also incur latency penalties for some of the iterations that are unrolled.

Looking at the architecture impact on throughput, a wider SIMD architecture produces a better throughput, but only when the application ILP allows it (see Tab. IV). This also means that more unrolling, which increases ILP also may better employ the architecture, thus increasing throughput. On the other hand, the pipeline length seems to have only marginal effect on the throughput for all techniques except, obviously, the single iteration schedule (see Tab. V).

### B. Code size

Modulo scheduling techniques fold the schedule into the *initiation interval*, that eventually is a fraction of a regular sequential schedule. As a result of this, modulo and unrolled schedule yield smaller code size compared to the overlapped execution, which basically includes all the iterations that are overlapped. Single iteration schedules are rather long in terms of code size, but are also sparse, including a number of *nop*s introduced due to dependencies and pipeline latency. Overall, modulo scheduling (which is equivalent to *unrolled x1*) is the most efficient when it comes to code size.

From an architectural point of view, wider SIMD units means also shorter code (see Tab. IV), when enough ILP is available in the application, since more work can be done in the same clock cycle. In contrast, longer pipelines seem to have again little influence on the code size for modulo and unrolled execution, but greatly affect the code size for overlapped and single iteration schedules. This is not unexpected, since the pipeline length directly dictates the schedule length for both of the latter, meanwhile modulo scheduling manages to wrap around (several times possibly) the length of the pipeline, hiding its latency.

### C. Storage requirements and data rates

Register pressure is decisive for architectures with small register files and slow memory. From Fig. 4 and all the tables, it is apparent that single execution always needs the least amount of registers, while overlapped execution uses the most, with modulo scheduling being somewhere in between. The amount of unrolling for modulo scheduling also directly affects the register pressure. Increasing the SIMD width, allows for more computations in parallel, thus increasing the pressure on the registers (see Tab. IV). Longer pipelines also seem to increase the register pressure, since data is alive over longer time intervals (see Tab. V).

When it comes to dimensioning buffers in streaming applications, input/output data rates are of paramount importance. Buffer sizes need to be selected in such a way that they never stall computations upstream or downstream due to lack of space or data. Opting for a high throughput implementation when the overall application context cannot support it would be a bad design choice. In here we refer to the distance between two consecutive inputs/outputs as *burstiness*. From Fig. 5, it is evident that overlapped execution is the most bursty scheduling technique while modulo is the least bursty. This effect is accentuated for longer pipelines. We note also that unrolling appears to increase the burstiness and the irregularity of input/output rates for modulo scheduling, proportionally to the number of iterations unrolled.

## VII. CONCLUSIONS

In this paper, we experimentally analysed three different scheduling techniques to compare their performance with respect to throughput, latency, register/memory pressure, data consumption/production rates and code size. We identified different bottlenecks caused by the application and the architecture. We then presented different trade-offs that are to be taken into account when employing one of the scheduling techniques, to compile a streaming application kernel on a generic parallel architecture that employs a SIMD hardware pipeline. Furthermore, we investigated the effects of altering the parameters of the generic architecture, namely the number of pipeline stages and SIMD width.